\documentclass[11pt]{article}             
\usepackage{osid}
\usepackage{graphics}
\usepackage{amssymb,amsmath}
\usepackage{rotating,curves}

\newcommand{\ket}[1]{|#1\rangle}
\newcommand{\bra}[1]{\langle #1|}

\title{Fidelity of Gaussian Channels}
\author{Carlton M.~Caves\thanks{$^{\ast}$Supported in part by US
    Office of Naval Research Grant No.~N00014-03-1-0426.}
    \\{\footnotesize\it Department of Physics and Astronomy,
    University of New Mexico, Albuquerque, NM~87131-1156, USA,
    email: caves@info.phys.unm.edu}\\[2ex]
        Krzysztof W\'odkiewicz\thanks{$^{\dagger}$Supported in part
    by KBN Grant No.~2PO3B\thinspace02123.}
    \\{\footnotesize\it Instytut Fizyki Teoretycznej, Uniwersytet Warszawski,
    Warszawa 00--681, Poland and Department of Physics and Astronomy,
    University of New Mexico, Albuquerque, NM~87131-1156, USA, email:
    wodkiew@fuw.edu.pl } }

\begin{document}

 \maketitle
\begin{abstract}
A noisy Gaussian channel is defined as a channel in which an input
field mode is subjected to random Gaussian displacements in phase
space.  We introduce the quantum fidelity of a Gaussian channel for
pure and mixed input states, and we derive a universal scaling law of
the fidelity for pure initial states.  We also find the maximum
fidelity of a Gaussian channel over all input states.  Quantum cloning
and continuous-variable teleportation are presented as physical
examples of Gaussian channels to which the fidelity results can be
applied.
\end{abstract}

\section{Introduction}

A principal aim of quantum information theory \cite{bennett1998} is to
determine the ultimate limits on capacity, entropy, or fidelity of
quantum information transmitted in noisy quantum channels
\cite{holevo1998,lloyd1997,hausladen1996,bennett1997,
adami1997,barnum1998}. Quantum channels with noise are examples of open
quantum systems that interact with an environment. The environment
produces classical and/or quantum fluctuations that generally degrade
the input states. The channel is described by a linear map $\rho
\mapsto \Phi(\rho)$, which takes an input state $\rho$ to an output
state $\Phi(\rho)$. The efficacy of the channel can be characterized by
a quantum fidelity $\mathcal{F}(\rho,\gamma)$, which measures the
quality of transmitted information as a function of the input state
$\rho$ and the noise parameter $\gamma$ that describes the environment.

A special class of noisy quantum channels consists of bosonic channels
with excess noise described by random Gaussian shifts in phase space
\cite{hall94,holevowerner2002,preskill2001,serafini2004}. Such quantum
Gaussian channels have attracted considerable attention in the the
framework of quantum information with continuous variables.  Recent
investigations devoted to bosonic Gaussian channels have addressed the
following problems: lower bounds on the capacities \cite{preskill2001},
the question of multiplicativity of the maximal $p$-norm output
purities \cite{serafini2004b}, and the Holevo-Werner additivity of
entropies at the output of the channels \cite{lloyd2004}.

The purpose of this paper is to describe the quantum fidelity of
Gaussian channels, using as a tool phase-space methods related to
Wigner and Weyl functions. A general scaling law for quantum fidelity
for pure input states is derived.  We argue that this scaling law
reflects the duality of the Wigner and Weyl phase-space distributions
related to sub-Planck \cite{zurek2001} and large scales of  the quantum
state $\rho$ (for an extended discussion of this duality, see
Ref.~\cite{Scott2004}).  We show that if the devices processing quantum
information can be built from linear optical elements in an extended
Hilbert space, the resulting Gaussian channels have important physical
applications. We discuss quantum cloning and a continuous-variable
teleportation protocol as examples of Gaussian noise channels.

The paper is structured as follows. In Sec.~\ref{sec:cpm} the
description of quantum channels in terms completely positive maps (CPM)
is recalled. In Sec.~\ref{sec:gauss} the Gaussian channel is defined in
terms of random Gaussian shifts and is related to Wigner and Weyl
functions in phase space. In Sec.~\ref{sec:fidelity} the quantum
fidelity of a Gaussian channel is introduced for pure and mixed input
states.  In Sec.~\ref{sec:scaling} a universal scaling law of the
fidelity for pure initial states is derived, and in
Sec.~\ref{sec:maxfidelity} we find the maximum fidelity of a Gaussian
channel over all input states. Section~\ref{sec:examples} presents as
examples explicit calculations of the fidelity for number states and
squeezed states.  Section~\ref{sec:ensemblefidelity} applies two
different kinds of fidelities for mixed input states to a thermal
input.  In Sec.~\ref{sec:physexamples} we discuss quantum cloning and
continuous-variable teleportation as physical examples of bosonic
Gaussian channels. Some final thoughts are presented in
Sec.~\ref{sec:conclusion}

\section{Quantum channels and completely positive maps}
\label{sec:cpm}

Consider a quantum system described by a Hilbert space $\mathcal{H}$,
with a given density operator $\rho$. A noisy quantum channel is a
linear and trace-preserving map of the quantum state $\rho$, given by
the Kraus decomposition
\begin{equation}
 \label{qchannel}
  \Phi(\rho) = \sum_i K_{i} \rho K^{\dagger}_i\,,
\end{equation}
where the completeness condition $ \sum_i K^\dagger_i K_i=I$ makes the
map trace-preserving \cite{kraus1983}. We call $\rho$ the input state
and the transformed $\Phi(\rho)$ the output state.

The overall system-environment state is described by a density operator
that evolves unitarily through the quantum channel. Most often, one is
interested in the system alone, which is described by a reduced density
operator obtained by tracing over the environment degrees of freedom.
As a result of such reduction, a quantum channel with noise is
characterized by a linear, trace-preserving, and completely positive
map.  A completely positive map is defined in the following way: if the
system undergoes the dynamics described by $\Phi$ and a reference
system $R$ of arbitrary Hilbert-space dimension experiences no
dynamics, a situation described by the overall superoperator
$\Phi\otimes\mathcal{I}_R$, where $\mathcal{I}_R$ is the identity
superoperator for the reference system, then an arbitrary (potentially
entangled) joint state of the system and reference system is mapped to
a positive output state; i.e., the superoperator $\Phi
\otimes\mathcal{I}_R$ maps positive operators to positive operators.

A completely positive map is required to describe reduced dynamics
because it implies that such dynamics arises from a unitary evolution
of the system and an environment,
\begin{equation}   \label{unitary}
    \Phi(\rho)=\textrm{Tr}_E [ U ( \rho \otimes
    | e\rangle \langle e|)U^\dagger ]\,.
\end{equation}
Here the environment degrees of freedom are denoted by $E$ and
$|e\rangle$ is some initial state of the environment.  The
following statements about a linear and trace-preserving map
$\Phi$ are equivalent: (i)~$\Phi$ is completely positive,
(ii)~$\Phi$ has a Kraus decomposition as in Eq.~(\ref{qchannel}),
and (iii)~$\Phi$ is the reduced dynamics for some
system-environment unitary dynamics as in Eq.~(\ref{unitary}).

\section{Bosonic Gaussian Channels}
\label{sec:gauss}

What we mean by a bosonic Gaussian channel in this paper is a
completely positive map that results from zero-mean random Gaussian
shifts in the phase space of a bosonic mode \cite{hall94}.  This CPM
acts on input states $\rho$ in the following way:
\begin{equation} \label{gaussdef}
\Phi(\rho) = \int d^2 \alpha\,
\mathcal{G}(\alpha)\, D(\alpha) \rho D^{\dagger}(\alpha)\,.
\end{equation}
Here $D(\alpha) =\exp(\alpha a^{\dagger} -\alpha^{\ast} a)$ is the
displacement operator in the phase space of the bosonic mode, where the
bosonic creation and annihilation operators obey the commutation
relation $[a,a^{\dagger}]=1$, and the classical-noise Gaussian
distribution,
\begin{equation}\label{gaussprofile}
 \mathcal{G}(\alpha)=\frac{2}{\pi \gamma}\,e^{-2|\alpha|^2/\gamma}\,,
\end{equation}
has zero mean and phase-space variance $\sigma^2=\gamma/2$. The complex
number $\alpha = (q+ip)/\sqrt2$ corresponds to the phase-space points
of a single-mode harmonic oscillator described by position $q$ and
momentum $p$. Notice that the Gaussian channel of Eq.~(\ref{gaussdef})
is a unital map; i.e., it leaves the identity operator unaffected. More
importantly, this Gaussian channel preserves the expectation value of
$a$; i.e., this kind of Gaussian channel cannot have any overall
attenuation or amplication.  The Gaussian channel (\ref{gaussdef}) is
in the Kraus form (\ref{qchannel}) if we make the identification $ K_i
\rightarrow \sqrt{\mathcal{G}(\alpha)}\,D(\alpha)$.

In Sec.~\ref{sec:physexamples} we show that such Gaussian channels can
be implemented in the framework of  linear optical elements in an
extended Hilbert space. In such cases the CPM~(\ref{gaussdef}) can be
derived from the general relation~(\ref{unitary}).  We will discuss
quantum cloning and continous-variable teleportation as physical
examples of bosonic Gaussian channels resulting from a reduction of
quantum systems that interact with an environment.

In our discussion of the Gaussian channel, we will analyze various
phase-space overlaps.  We use the following definitions of the
phase-space Wigner and Weyl functions. The Wigner function of an
arbitrary density operator $\rho$ is \cite{garzoll} given by
\begin{equation}  \label{wigdef}
  W_{\rho}(\alpha) = \frac{2}{\pi} \mathrm{Tr}\!\Bigl[\,
\rho D(\alpha) (-1)^{a^{\dagger} a} D^\dagger(\alpha)\Bigr]
= \int
\frac{d^2\beta}{\pi^2} \, e^{\alpha
\beta^{\ast}-\alpha^{\ast}\beta}\,C_{\rho}(\beta)\,,
\end{equation}
where $C_{\rho}(\alpha) =\mathrm{Tr}[ \rho D(\alpha)]$ is the Weyl
characteristic function and $\rho$ is an arbitrary density operator. We
find it useful to rewrite the Gaussian CPM (\ref{gaussdef}) in terms of
the phase-space Wigner functions of the input and the output states:
\begin{equation} \label{gausswigdef}
W_{\Phi(\rho)}(\alpha) =\int d^2 \beta\,
\mathcal{G}(\alpha-\beta)\, W_\rho(\beta)\,.
\end{equation}
The counterpart of this relation in terms of the Weyl functions of the
input and output states is
\begin{equation}  \label{gaussweyldef}
C_{\Phi(\rho)}(\alpha)=e^{-\gamma|\alpha|^2/2}C_\rho\,.
\end{equation}

\section{Fidelity of a Gaussian channel}
\label{sec:fidelity}

\subsection{Fidelity for pure input states}

If the input state to the Gaussian channel is a pure state $\rho =
|\Psi \rangle\langle  \Psi |$, the channel fidelity is given as a
quantum overlap between the input and the output state.  In this case
the channel fidelity is
\begin{equation}
\label{purefidel}
\mathcal{F}(\Psi,\gamma)= \bra{\Psi}\Phi(\rho)\ket{\Psi}\,.
\end{equation}
Using the phase-space Wigner functions of the input and output state,
we can rewrite this fidelity as a phase-space overlap
\begin{equation}  \label{wigfidel}
 \mathcal{F}(\Psi,\gamma)= \pi \int d^2\alpha
\, W_{\Phi(\rho)}(\alpha) W_{\Psi}(\alpha)
=\pi\int d^2\alpha\,d^2\beta\,\mathcal{G}(\alpha-\beta)
W_\Psi(\alpha)W_\Psi(\beta)\,.
\end{equation}
Another useful form for this fidelity, this time in terms of the Weyl
functions, comes from substituting the output state~(\ref{gaussdef})
directly into Eq.~(\ref{purefidel}):
\begin{equation}  \label{weylfidel}
\mathcal{F}(\Psi,\gamma)=
\int d^2\alpha\,\mathcal{G}(\alpha)|C_\Psi(\alpha)|^2\,.
\end{equation}

\subsection{Fidelity for mixed input states}
\label{sec:fidelitymixed}

An appropriate measure for assessing the fidelity of an mixed input
state is the entanglement fidelity \cite{schumacher1996}, which is
defined in the following way.  Imagine that the input mixed state
$\rho$ is purified to a state $|\psi\rangle$ of the original mode and a
reference mode $R$.  An example of such a purification is
\begin{equation}  \label{purif}
 |\psi\rangle = \sqrt{\rho} \otimes I_R \sum_n\,|n\rangle\,
 \otimes |n\rangle
 = \sum_n\sqrt\rho\,|n\rangle\otimes|n\rangle\,.
\end{equation}
The input mode is now entangled with the reference mode, and the
purified state belongs to an enlarged Hilbert space
$\mathcal{H}\otimes\mathcal{H}_R$.  We now suppose that the original
mode is subjected to the Gaussian channel while the reference mode is
left untouched.  The resulting output state is
\begin{equation}
\Phi\otimes\mathcal{I}_R(|\psi\rangle\langle\psi|)=
\int d^2 \alpha\,\mathcal{G}(\alpha)\,
D(\alpha)\otimes I_R\,|\psi\rangle\langle\psi|\,I_R\otimes D^{\dagger}(\alpha)\,.
\end{equation}
The entanglement fidelity is now defined to be the fidelity of this
joint output state with the purified input state:
\begin{equation}
\mathcal{F}(\rho,\gamma)=
\bigl\langle\psi\bigl|
\Phi\otimes\mathcal{I}_R(|\psi\rangle\langle\psi|)\bigr|\psi\bigr\rangle\,.
\end{equation}
The entanglement fidelity is a property of the system state $\rho$ even
though the purification is not unique.  The entanglement fidelity
reduces to the fidelity of Eq.~(\ref{purefidel}) in the case of pure
input states.

It is now easy to see that the entanglement fidelity for a Gaussian
channel has the form
\begin{equation}  \label{gaussentfidel}
\mathcal{F}(\rho,\gamma)=
\int d^2\alpha\,\mathcal{G}(\alpha)|C_\rho(\alpha)|^2\,,
\end{equation}
which is the same as the corresponding form~(\ref{weylfidel}) of the
pure-state fidelity.  Notice, however, that the entanglement fidelity
is not given by the mixed-state version of Eq.~(\ref{wigfidel}).  A
more thorough discussion of the entanglement fidelity for mixed input
states is given in Ref.~\cite{Scott2004}.

Another possible fidelity measure for a mixed-state input regards the
mixed state as coming from a particular ensemble of states,
$\{p_n,\Psi_n\}$, by which we mean that in each experimental run, one
of the states $\ket{\Psi_n}$ is selected randomly with probability
$p_n$. Thus, in each run, the Gaussian channel delivers with
probability $p_n$ the initial state with fidelity
$\mathcal{F}(\Psi_n,\gamma)$. A statistical average leads to a mean
ensemble fidelity given by
\begin{equation}
\label{ensfidel}
\bar{\mathcal{F}}(\{p_n,\Psi_n\},\gamma)=
\sum_n p_n \mathcal{F}(\Psi_n,\gamma)
=\sum_n p_n \bigl\langle\Psi_n\bigl|
\Phi\bigl(|\Psi_n\rangle\langle\Psi_n|\bigr)\bigr|\Psi_n\bigr\rangle \,.
\end{equation}
The mean ensemble fidelity depends on the particular ensemble used to
make up an input density operator $\rho=\sum_n
p_n|\Psi_n\rangle\langle\Psi_n|$.  The convexity of the entanglement
fidelity means that the entanglement fidelity for a mixed input state
$\rho$ is less than or equal to the mean ensemble fidelity for any
ensemble corresponding to $\rho$.

\section{Scaling law of channel fidelity}
\label{sec:scaling}

If we assume that the input state is given by a pure state, the
fidelity is given by the quantum overlap formula~(\ref{purefidel}),
which leads to the equivalent phase-space formulas~(\ref{wigfidel}) and
(\ref{weylfidel}).  Here we use these phase-space formulas to show that
the fidelity of the Gaussian channel obeys a universal scaling law.

Applying the Fourier transform relation~(\ref{wigdef}) between the
Wigner and Weyl functions to Eqs.~(\ref{wigfidel}) and
(\ref{weylfidel}), we obtain two new formulas for the fidelity.  The
result is four equivalent forms for the channel fidelity:
\begin{eqnarray}
\label{fideloverlap}
 \mathcal{F}(\Psi,\gamma) &=&\frac{2}{\pi
\gamma}\int d^2 \alpha\,
e^{-2|\alpha|^2/\gamma}|C_{\Psi}(\alpha)|^2 \nonumber\\
&=&\frac{2}{ \gamma}\int d^2 \alpha \,d^2 \beta\,
e^{-2|\alpha -\beta|^2/\gamma}
W_{\Psi}(\alpha)W_{\Psi}(\beta)\nonumber\\
&=&\frac{1}{\pi}\int d^2 \alpha\,
e^{-\gamma|\alpha|^2/2}|C_{\Psi}(\alpha)|^2 \nonumber\\
 &=&\int d^2
\alpha \,d^2 \beta\, e^{-\gamma|\alpha -\beta|^2/2}
W_{\Psi}(\alpha)W_{\Psi}(\beta)\,.
\end{eqnarray}
The first two lines are rewrites of Eqs.~(\ref{wigfidel}) and
(\ref{weylfidel}).  The third line comes from writing the
fidelity~(\ref{purefidel}) as an overlap of the input and output Weyl
characteristic functions; it is thus also obtained by Fourier
transforming the Wigner functions in the integrand of the second line.
Similarly, the last line is obtained by Fourier transforming the Weyl
functions in the integrand of the first line.

The first and third forms (and the second and fourth) show us that
\begin{equation}   \label{scaling}
\mathcal{F}(\Psi,\gamma) =
\frac{2}{\gamma}\mathcal{F}(\Psi,4/\gamma)\,.
\end{equation}
This scaling law reflects a duality between Wigner functions and Weyl
characteristic functions.  This duality is related to sub-Planck
structures of the Wigner functions~\cite{zurek2001}. For a given input
state, the Wigner function has two important scales: a small scale $l$
and a large scale $L$.  The small scale $l$ characterizes the
sub-Planck phase-space structures in the input state's Wigner function.
The large scale $L$ characterizes the scale over which the Wigner
function is nonnegligible. For pure states, these two scales are
related by an uncertainty relation, $L l \sim 1$. Since the Weyl
characteristic function is the Fourier transform of the Wigner
function, these two scales appear inversely in the Weyl function.

The scaling law~(\ref{scaling}) displays this duality. If we look at
the fidelity in the form of the second relation in
Eq.~(\ref{fideloverlap}), we see that the fidelity between the input
state and the output state approaches unity if the dispersion of the
Gaussian channel satisfies $\sigma=\sqrt{\gamma/2} \le l$ so that the
integral approaches ${\rm Tr}(\rho^2)$, which is 1 for a pure state; in
this case, all the small-scale phase-space structure of the Wigner
function is well transmitted through the noisy channel.  The dual form
of the fidelity, expressed in the the last relation of
Eq.~(\ref{fideloverlap}), says that to get good fidelity, we need to
have $1/\sigma=\sqrt{2/\gamma}\ge L$, so that the integral reduces to
the square of the integral over the entire Wigner function.  Putting
these two results together gives the phase-space uncertainty relation,
$Ll\sim 1$.

In terms of the dispersion, the scaling law has the form
\begin{equation}
\mathcal{F}(\Psi,\sigma) =
\frac{1}{\sigma^2}\mathcal{F}(\Psi,1/\sigma)\,.
\end{equation}
Additional discussion of the relation between fidelity and sub-Planck
structure and of the phase-space uncertainty relation can be found in
Ref.~\cite{Scott2004}.

\section{Maximum fidelity}
\label{sec:maxfidelity}

From these considerations, we can derive the  maximum channel
fidelity that can be achieved by any initial pure state. We show
that the maximum fidelity is given by the coherent-state fidelity
for all values of $\gamma$. To demonstrate this, return to the
expression for the fidelity given by the last formula in
Eq.~(\ref{fideloverlap}):
\begin{equation}
 \mathcal{F}(\Psi ,\gamma)=
\int d^2\alpha\, d^2\beta\,
e^{-\gamma|\alpha-\beta|^2/2} W_\Psi (\alpha)W_\Psi
(\beta)\,.
\end{equation}
The task can be restated as finding the pure state that maximizes
this overlap.

Notice that this fidelity can be thought of as the average value of
$e^{-\gamma|\alpha-\beta|^2/2}$ with respect to a pure product copy
state, $|\Psi_A\rangle\otimes|\Psi_B\rangle$, of two modes, $A$ and
$B$; the joint Wigner function of the two modes is
$W_{AB}(\alpha,\beta)=W_\Psi(\alpha )W_\Psi(\beta)$. Introducing modes
$C$ and $D$, with annihilation operators $c=(a+b)/\sqrt{2}$ and
$d=(a-b)/\sqrt{2}$ and corresponding c-number variables
$\chi=(\alpha+\beta)/\sqrt{2}$ and $\delta=(\alpha-\beta)/\sqrt{2}$, we
can rewrite the fidelity as
\begin{equation}
\mathcal{F}(\Psi ,\gamma)=\int d^2
\delta\,e^{-\gamma|\delta|^2}W_D(\delta)\,,
\end{equation}
where $W_{D}(\delta)=\int d^2\chi \,W_{AB}(\chi,\delta)$. What we see
is that the  fidelity is the expectation value of the mode-$D$ operator
$A_\gamma$ whose symmetrically ordered associated function is
$e^{-\gamma|\delta|^2}$.  Letting $\gamma=(\bar{n}+1/2)^{-1}$, we see
that $A_\gamma$ is given by $\bar{n}+1/2$ times the density operator
for a thermal state of mode $D$ whose mean number of photons is
$\bar{n}=\gamma^{-1}(1-\gamma/2)$. Thus we can write the fidelity as
\begin{equation}
 \mathcal{F}(\Psi,\gamma)= {\rm Tr}(A_\gamma\rho_D)\,,
\end{equation}
where
\begin{eqnarray}
A_\gamma &=&\frac{1/2+\bar{n}}{1+\bar{n}}
\left(\frac{\bar{n}}{1+ \bar{n}}\right)^{d^\dagger d} \nonumber \\
&=&\frac{1}{1+\gamma/2}
\left(\frac{1-\gamma/2}{1+\gamma/2}\right)^{d^\dagger d} \nonumber\\
&=& \frac{1}{1+\gamma/2}
\left(\frac{1-\gamma/2}{1+\gamma/2}\right)^{(a^\dagger-b^\dagger)(a-b)/2}\,.
\end{eqnarray}

Notice that $A_{\gamma=0}=I_D$, confirming that the fidelity is 1
regardless of the input state for $\gamma=0$. Generally we can bound
the fidelity by the largest eigenvalue of $A_\gamma$:
\begin{equation}  \label{boundfidel}
\mathcal{F}(\Psi,\gamma)\leq
\Bigl(\mbox{max eigenvalue of $A_\gamma$}\Bigr)
=\frac{1}{1+\gamma/2}\,.
\end{equation}
The reason this is the largest eigenvalue is that the factor in large
parentheses in the expression for $A_\gamma$ has magnitude $\le1$,
which means that the largest eigenvalue, corresponding to the vacuum
state for mode~$D$, is $(1+\gamma/2)^{-1}$.  Since coherent states
saturate the upper bound, we can write
\begin{equation}  \label{maxfidel}
\mathcal{F}_{\rm max}(\gamma)=\frac{1}{1+\gamma/2}\,.
\end{equation}

The bound on the expectation value of $A_\gamma$ is useful in other
applications than the noisy Gaussian channels considered here.  For
that purpose, note that the bound holds for all joint states
$\rho_{AB}$ of modes $A$ and $B$---i.e., it holds for the expectation
value ${\rm Tr}(A_\gamma\rho_{AB})$---not just for the pure product
copy states that are relevant to the channel fidelity.  A joint state
achieves the expectation value bound if and only if mode $D$ is in
vacuum; i.e., the state $\rho_{AB}$ is the state of the two output
modes, $a=(c+d)/\sqrt{2}$ and $b=(c-d)/\sqrt{2}$, of a 50:50
beamsplitter that has vacuum incident on its mode-$D$ input.

If we specialize to pure product input states,
$|\Psi_A\rangle\otimes|\Phi_B\rangle$, we can say much more about when
the bound is achieved.  Since mode~$D$ is in vacuum, we have
\begin{equation}
0=d|\Psi_A\rangle\otimes|\Phi_B\rangle
=\frac{1}{\sqrt2}(a-b)|\Psi_A\rangle\otimes|\Phi_B\rangle\,,
\end{equation}
which implies that
\begin{equation}
 a|\Psi_A\rangle\otimes|\Phi_B\rangle=
|\Psi_A\rangle\otimes\,b|\Phi_B\rangle\,.
\end{equation}
This requires that $a|\Psi_A
\rangle=\langle\Phi_B|b|\Phi_B\rangle|\Psi_A\rangle=\alpha|\Psi_A
\rangle$ and
$b|\Phi_B\rangle=\langle\Psi_A|a|\Psi_A\rangle|\Phi_B\rangle=\alpha
|\Phi_B \rangle$, i.e., that $|\Psi_A\rangle$ and $|\Phi_B\rangle$ are
the same coherent state $|\alpha\rangle$.  Thus the maximum channel
fidelity is achieved if and only if the input state is a coherent
state.

\section{Examples of pure states in a Gaussian channel}
\label{sec:examples}

\subsection{Number states}

Let us take as an example the case of an input state that is a number
state $|n\rangle$ of a harmonic oscillator. In this case the fidelity
of transmission through a Gaussian channel is
\begin{equation}
\mathcal{F}(|n\rangle,\gamma) =
\frac{2}{\pi \gamma}
\int d^{2} \alpha\,e^{-2|\alpha|^{2}/\gamma}
|\langle n|D(\alpha)|n\rangle|^{2}\,.
\end{equation}
Using the property
\begin{equation}
\langle n|D(\alpha)|n\rangle = e^{-|\alpha|^{2}/2}L_{n}(|\alpha|^{2})\,,
\end{equation}
where $L_{n}$ denotes the $n$th-order Laguerre polynomial, one can
calculate an exact expression for the fidelity generating function for
all number states:
\begin{eqnarray}
 \mathcal{F}(\gamma,\lambda)=\sum_{n=0}^{\infty}
\lambda^{n}\mathcal{F}(|n\rangle,\gamma)&=&
\frac{1}{\sqrt{(1+\gamma/2)^{2}-2\lambda(1+\gamma^{2}/4)
+\lambda^{2}(1-\gamma/2)^{2}}}
\nonumber \\
&=&
\frac{1}{\sqrt{[(1-\lambda)+(1+\lambda)\gamma/2]^2-\lambda\gamma^2}}
\,.
\label{generfun}
\end{eqnarray}
The resulting fidelity of a Gaussian channel with a number state
at the input is
\begin{equation}  \label{nLegendre}
\mathcal{F}(|n\rangle,\gamma)=
\frac{(1-\gamma/2)^n}{(1+\gamma/2)^{n+1}}\,
P_n\!\left(\frac{1+\gamma^2/4}{1-\gamma^2/4}\right)\;,
\end{equation}
where $P_n(x)$ is a Legendre polynomial.  For $\gamma=1,2$, this
becomes
\begin{equation}
\mathcal{F}(|n\rangle,1) = \frac{2}{3^{n+1}}P_n(5/3)\,,\quad \quad
\mathcal{F}(|n\rangle,2) = \frac{(2n)!}{2^{2n+1}(n!)^2}\,.
\end{equation}
Using a series expansion of  the generating function~(\ref{generfun})
or working directly with the expression~(\ref{nLegendre}), one can
easily calculate the fidelities for the lowest number states:
\begin{equation}
\mathcal{F}(|1\rangle,\gamma) =
2\frac{4+\gamma^2}{(2+\gamma)^3}\,,\quad \quad
\mathcal{F}(|2\rangle,\gamma)
=2\frac{16+16\gamma^2+\gamma^4}{(2+\gamma)^5}\,.
\end{equation}

Techniques similar to those used in this section can be used to
calculate the channel fidelity for an input state that is an arbitrary
superposition of number states, but in the absence of some general
technique like the generating function~(\ref{generfun}), the
calculation becomes increasingly tedious as higher number states are
included in the superposition.  As an example, for input state $|\Psi
\rangle=(|0\rangle+|1\rangle)/\sqrt2$, the channel fidelity is
\begin{equation}
 \mathcal{F}(\Psi,\gamma)=
\frac{1+3\gamma/4+\gamma^2/4}{(1+\gamma/2)^3}\,.
\end{equation}

\subsection{Squeezed state}

A squeezed state of a one-dimensional harmonic oscillator is given
by the following formula \cite{schumaker}:
\begin{equation}
\ket{\mu} =
(1-|\mu|^2)^{1/4}\,e^{-\mu{a^{\dagger}}^2/2}|0\rangle \,.
\end{equation}
Here
\begin{equation}
|\mu|=\tanh r=\sqrt{\frac{\bar{n}}{1+\bar n}}\,,
\end{equation}
where $r$ is the usual squeeze parameter and $\bar n$ is the mean
number of oscillator quanta.  For $\mu=0$, $\bar{n}=0$, and the
squeezed state reduces to the ground state of the one-dimensional
harmonic oscillator.  The fidelity of a Gaussian channel with a
squeezed state at the input is
\begin{equation}  \label{squeezedfidel}
\mathcal{F}(|\mu\rangle,\gamma)=
\frac{1}{\sqrt{1+(2\bar n+1)\gamma+\gamma^2/4}}\,.
\end{equation}

Figure~1 depicts the fidelity of various states discussed in this
section, as functions of the channel noise $\gamma$.

\begin{figure}[h]
\begin{center}
  \includegraphics[height=9cm, width= 9cm]{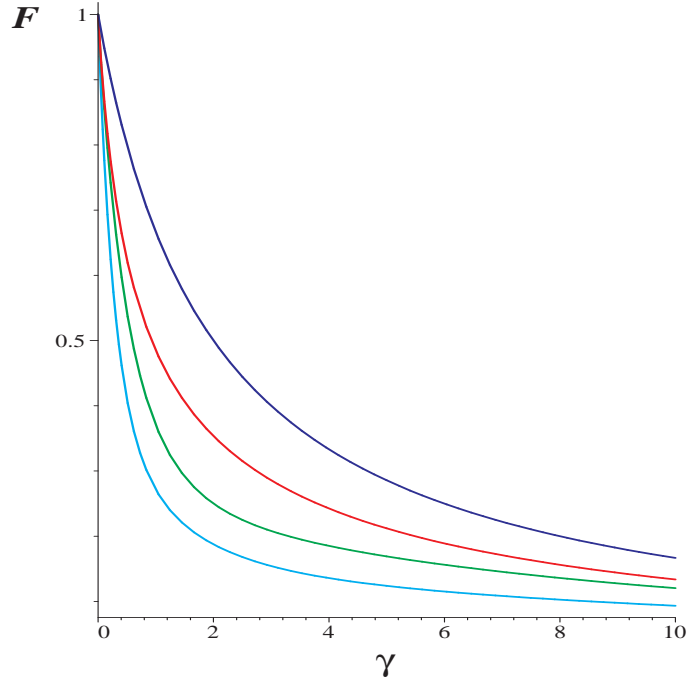}\\
  \caption{Plots of the channel fidelity as functions of $\gamma$ for
  different input states. From the top: the vacuum state $|0\rangle$,
  the squeezed state $|\mu\rangle$ with $\bar{n}=1$, and the first two number states,
  $|1\rangle$ and $|2\rangle$.}
  \end{center}
\end{figure}

\section{Example of mixed-state fidelities for a Gaussian channel}
\label{sec:ensemblefidelity}

Because of the many experimentally uncontrollable properties of the
input states to a quantum channel, in reality we have to deal with
sources described by a mixed state or, in some cases, by a particular
statistical ensemble of incoming pure states.  In
Sec.~\ref{sec:fidelitymixed} we introduced the entanglement fidelity
and the mean ensemble fidelity as fidelity measures to characterize
these situations.  We note that since the mean ensemble
fidelity~(\ref{ensfidel}) is an average of pure-state fidelities, it
satisfies the scaling law~(\ref{scaling}) and the upper
bound~(\ref{boundfidel}). Moreover, since the entanglement fidelity is
bounded above by the mean ensemble fidelity, we can write generally
that
\begin{equation}
\mathcal{F}(\rho,\gamma)\le
\bar{\mathcal{F}}(\{p_n,\Psi_n\},\gamma)
\leq \frac{1}{1+\gamma/2}
\end{equation}
for any ensemble that corresponds to the input state $\rho$.

In this section we illustrate the two mixed-state fidelities by
considering an ensemble of number states, $\{p_n,\ket{n}\}$, that are
selected with the Bose-Einstein probabilities
\begin{equation}
p_n = \frac{\bar{n}^n}{(1+\bar{n})^{n+1}}\,.
\end{equation}
The corresponding density operator is the thermal density operator
\begin{equation}
\rho=\sum_{n=0}^\infty p_n|n\rangle\langle n|=
\frac{1}{1+\bar n}\left(\frac{\bar n}{1+\bar n}\right)^{a^\dagger a}\,,
\end{equation}
where $\bar n$ is the mean number of thermal quanta.

\begin{figure}[h]
\begin{center}.
  \includegraphics[height=9cm, width= 9cm]{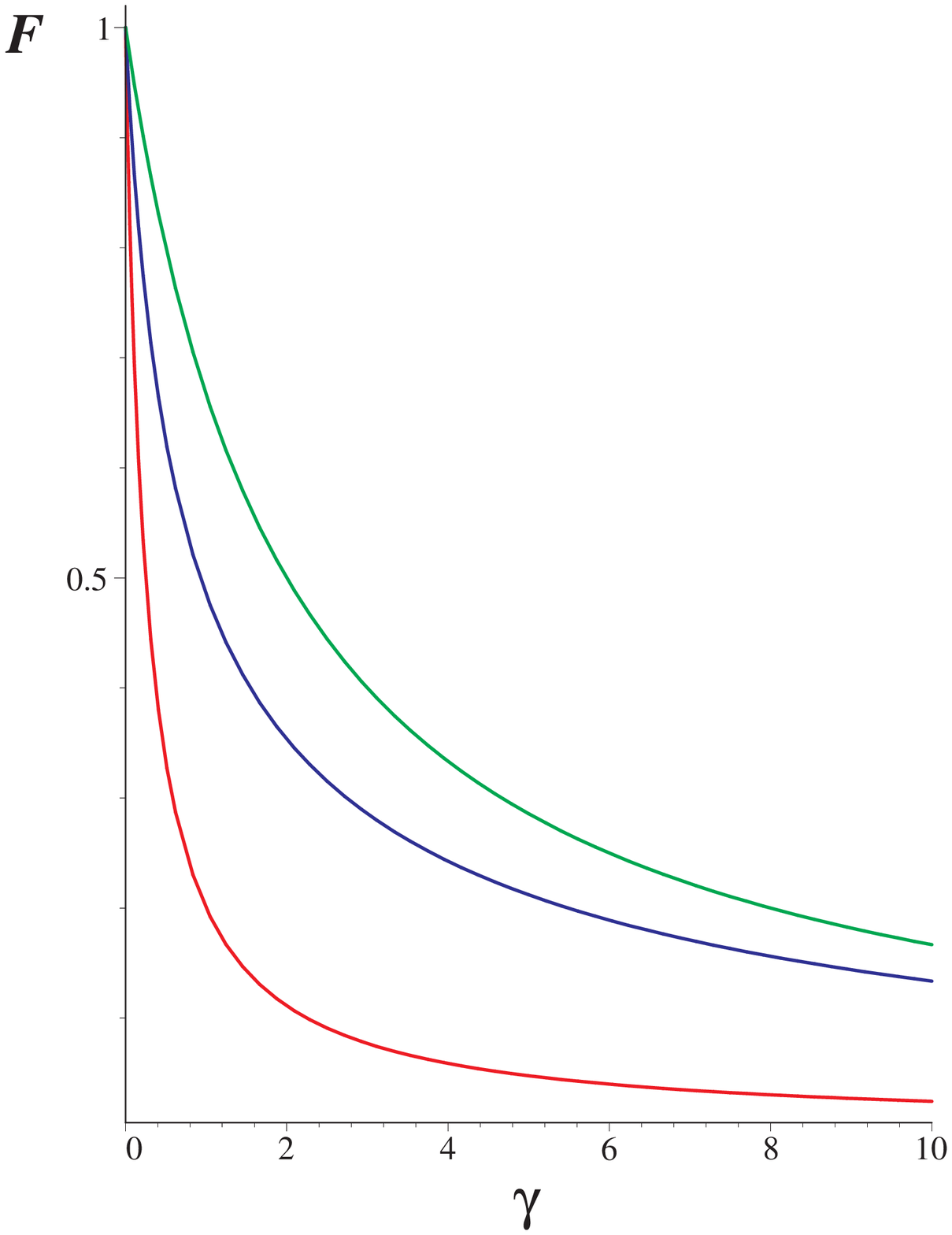}\\
  \caption{Plots of the entanglement fidelity (lower curve) and the
  mean ensemble fidelity (middle curve) as functions of $\gamma$ for
  $\bar{n}=1$. For reference, the upper curve gives the fidelity of a
  coherent state.}
  \end{center}
\end{figure}

In this case we can apply the formula~(\ref{generfun}) to find the mean
ensemble fidelity.  A simple calculation gives
\begin{equation}  \label{thermalmeanF}
\bar{\mathcal{F}}=
\frac{1}
{\sqrt{[1+(2\bar n+1)\gamma/2]^2-\bar n(1+\bar n)\gamma^2}}
=\frac{1}{\sqrt{1+(2\bar n+1)\gamma+\gamma^2/4}}\,,
\end{equation}
the same as the fidelity~(\ref{squeezedfidel}) for an input squeezed
state with the same mean number of quanta. It is easy to verify that
this ensemble fidelity satisfies the scaling law~(\ref{scaling}). The
entanglement fidelity follows from inserting the Weyl function for a
thermal state, $C_\rho(\alpha)=e^{-|\alpha|^2(\bar n+1/2)}$, into
Eq.~(\ref{gaussentfidel}), which gives
\begin{equation}  \label{thermalentF}
\mathcal{F}(\rho,\gamma)=\frac{1}{1+(2\bar n+1)\gamma/2}\,.
\end{equation}
It is trivial to see in this case that $\mathcal{F}(\rho,\gamma)\le
\bar{\mathcal{F}}$, with equality holding if and only if $\bar n=0$ or
$\gamma=0$. In Fig.~2 we have plotted, as functions of the channel
noise $\gamma$, the mean ensemble fidelity and the entanglement fidelity
for an ensemble with Bose-Einstein statistics.

\section{Physical examples of Gaussian channels}
\label{sec:physexamples}

\subsection{Cloning}

As an initial example of a Gaussian channel, we now investigate a very
simple experimental setup used for approximate cloning of the states of
a field mode.  The no-cloning theorem \cite{wooters,yuen,barnum} shows
that a universal and faithful cloning machine, which would clone an
arbitrary input quantum state perfectly, is incompatible with quantum
mechanics. It is possible, however, to find imperfect cloning machines
that copy quantum states with some loss of quantum fidelity. The
simplest device designed to clone quantum states is a 50:50 beam
splitter that is preceded by an amplifier with amplitude gain of
$\sqrt2$ to compensate for the reduction in signal amplitude at the
beam splitter.  This setup is depicted in Fig.~3.

\begin{figure}
\begin{center}
\includegraphics[scale=0.7]{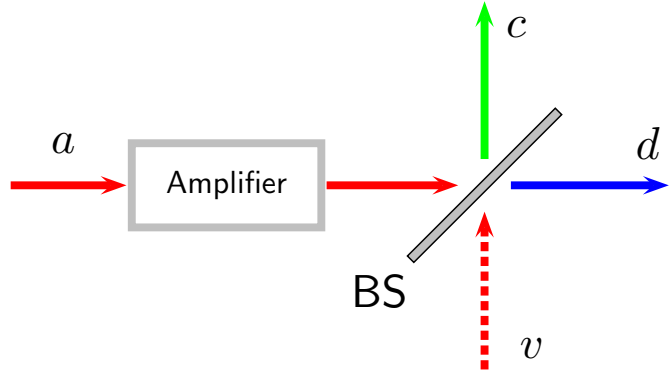}
  \caption{Cloning experimental setup with an amplifier of amplitude
gain $\sqrt{2}$.  BS denotes a 50:50 beam splitter.  The annihilation
operators of the input mode, the vacuum mode, and the modes that carry
the clones are indicated.}
\end{center}
\end{figure}

The input (pure) state of mode $A$ is cloned into two outgoing clones
in modes $C$ and $D$.  The setup of this cloning device requires only
linear optics and linear amplification.  As a result, the annihilation
operators for the outgoing modes of the clones are \cite{caves82}
\begin{eqnarray}
 c &=&\frac{1}{\sqrt{2}}(\sqrt2 a+a_{\rm amp}^\dagger+v)
 =a+\frac{1}{\sqrt2}(a_{\rm amp}^\dagger+v)\,,\nonumber \\
 d &=&\frac{1}{\sqrt{2}}(\sqrt2 a+a_{\rm amp}^\dagger-v)
 =a+\frac{1}{\sqrt2}(a_{\rm amp}^\dagger-v)\,,
\end{eqnarray}
where $a_{\mathrm{amp}}^\dagger$ is a vacuum-noise creation operator
that describes the noise introduced by the amplifier, and $v$ is the
annihilation operator for the vacuum-mode $V$ that is incident on the
unused port of the beam splitter.  The whole dynamics of the cloning
process, including amplification, is described by a unitary
transformation in the extended Hilbert space
$\mathcal{H}_A\otimes\mathcal{H}_{\mathrm{\mathrm{amp}}}\otimes
\mathcal{H}_V$ of the input mode $A$, the amplifier-noise mode, and the
vacuum mode $V$.  As a result of this transformation, the reduced
dynamics of such a cloning device corresponds to a Gaussian channel
with $\gamma=1$.

For all initial pure states, the cloning fidelity is given by
\begin{equation}
\mathcal{F}_{\mathrm{clone}}(\Psi)=
\mathcal{F}(\Psi,\gamma=1)=
2\mathcal{F}(\Psi,4)\,,
\end{equation}
where the latter equality is a consequence of the scaling
law~(\ref{scaling}). The general fidelity bound~(\ref{boundfidel}) for
a Gaussian channel gives us a bound on the cloning fidelity,
$\mathcal{F}_{\mathrm{clone}}\le\mathcal{F}_{\rm max}(\gamma=1)=\frac{2}{3}$,
with equality achieved only for cloning of coherent states.

\subsection{Continuous-variable teleportation}

As a second example, we consider a continuous-variable quantum
teleportation protocol, which is effectively equivalent to a Gaussian
channel~\cite{cmckw}. Quantum teleportation is a process that can
transfer an arbitrary quantum state from a system held by one party,
usually called Alice, to a system held by a second party, usually
called Bob \cite{bennett,vaidman,braunsteinkimble}. The process
requires a pair of systems, shared by Alice and Bob, in an entangled
state---the entangled resource---and an amount of classical information
transmitted from Alice to Bob.

The entangled resource is ideally a pure two-mode squeezed state
\cite{schumaker} of two modes, $A$ and $B$, which have annihilation
operators $a$ and $b$.  Such states can be described by a Gaussian
Wigner function, $W_{AB}(\alpha,\beta)$, that is specified by the
following nonvanishing second moments:
\begin{eqnarray}
n+\frac{1}{2} &=&  \int d^2\alpha\, d^2\beta \, |\alpha|^2
W_{AB}(\alpha,\beta) =\int d^2\alpha\, d^2\beta\,
|\beta|^2\,W_{A,B}(\alpha,\beta)\,,\nonumber\\
m &=&  \int d^2\alpha\, d^2\beta \,\alpha\beta\,
W_{AB}(\alpha,\beta)
=\int d^2\alpha\, d^2\beta \,\alpha^*\beta^*\,
W_{AB}(\alpha,\beta)\,.
\label{squeezedmoments}
\end{eqnarray}
The moments must satisfy $n\ge0$ and $\sqrt{n(n+1)}\ge|m|$ to
ensure that the Wigner distribution corresponds to a valid quantum
state.  The state is pure if and only if $\sqrt{n(n+1)}=|m|$, in
which case it is a two-mode squeezed state.  When the state is
pure, the limit $|m|\rightarrow\infty$ gives the original
entangled state of Einstein, Podolsky, and Rosen~\cite{EPR35},
with $m$ negative leading to a Wigner function proportional to
$\delta(q_A+q_B)\delta(p_A-p_B)$, which are the ideal correlations
for the teleporation protocol we are considering, and $m$ positive
leading to a Wigner function proportional to
$\delta(q_A-q_B)\delta(p_A+p_B)$.  The correlated state of
Eq.~(\ref{squeezedmoments}) is separable (unentangled) if and only
if $n\geq|m|$ \cite{tutorial}.

The state to be teleported is an input pure state $\rho$ of a mode $V$
in Alice's possession, which has annihilation operator $v$. The
protocol consists of (i)~Alice's measuring the two (commuting) homodyne
quadratures contained in the Hermitian real and imaginary parts of the
operator $v+a^{\dagger}$; (ii)~Alice's communicating to Bob the
(complex) result $\xi$ of this measurement; and (iii)~Bob's displacing
the complex amplitude of his mode $B$ by $\xi$.

The efficacy of the protocol is quantified by the fidelity between the
output state of mode $B$ and the input state $|\psi\rangle$, averaged
over the possible measurement results.  The teleportation protocol
involves three modes in an extended Hilbert space
$\mathcal{H}_V\otimes\mathcal{H}_A\otimes\mathcal{H}_B$. Reduction of
this protocol to the Hilbert space of the incoming mode reduces the
teleportation protocol to a Gaussian channel with quantum fidelity
given by Eq.~(\ref{wigfidel}), where the Gaussian noise distribution is
calculated from the following relation:
\begin{equation}
G(\nu)= \int d^2\alpha\,d^2\beta\,
\delta(\beta+\alpha^*-\nu)W_{AB}(\alpha,\beta)\,.
\end{equation}
Simple calculation involving the Wigner function of the entangled
resource leads to a Gaussian distribution for $G(\nu)$, with the noise
parameter given by $\gamma = 2[1+2(n+m)]$. A separable resource has
$n\geq|m|$ and, accordingly, $\gamma\geq2$.  For $\sqrt{n(n+1)}\geq
m\ge n$, the correlated state is entangled, but with the wrong sort of
correlations for the protocol we are considering, so $\gamma\ge 2$. For
$\sqrt{n(n+1)}\geq-m\ge n$, the correlated state is entangled and
$0\le\gamma\le2$. Perfect teleportation is achieved if $\gamma=0$,
which corresponds to $m=-\sqrt{n(n+1)}\rightarrow-\infty$ ($m+n=-1/2$),
i.e., to a pure entangled state with perfect EPR correlations.  For
pure input states, we conclude that the maximum teleportation fidelity
for a given entangled resource $\gamma$ is
\begin{equation}
 \mathcal{F}_{\rm max}(\gamma)=
 \frac{1}{1+\gamma/2}=\frac{1}{2(1+ n+m)}
\end{equation}
and is achieved if and only if the input state is a coherent state.

We have used the fidelity bound~(\ref{maxfidel}) to get two other
interesting results \cite{cmckw}.  The first is that the maximum
fidelity for teleporting a coherent state using the standard protocol,
but with any separable state for modes $A$ and $B$ [not necessarily a
state of the Gaussian form specified by the
moments~(\ref{squeezedmoments})], is $1/2$. The second result has to do
with local hidden-variable models for continuous-variable
teleportation. The teleportation of any Gaussian input state can be
described within a local hidden-variable model, no matter what fidelity
is achieved in the teleportation; the hidden-variable model is based on
the classical phase-space variables of the Wigner distribution.  For
non-Gaussian pure input states, we have shown that the value $\gamma=1$
plays a special role: each non-Gaussian pure input state $|\Psi\rangle$
has its own threshold fidelity, ${\cal F}(\Psi,1)<{\cal F}_{\rm
max}(\gamma=1)=2/3$, below which its teleportation can be accommodated
within an extended phase-space hidden-variable model and above which it
cannot.

\section{Conclusion}
\label{sec:conclusion}

In this paper we follow Hall~\cite{hall94} in defining a Gaussian
channel as one in which an input mode is subjected to random Gaussian
displacements in phase space.  Such channels arise naturally whenever
field modes undergo linear optical transformations, linear
amplification, and measurements of quadrature components, provided that
the overall channel preserves the mean complex amplitude of the input
mode.  We introduce the quantum fidelity for both pure and mixed inputs
and derive the maximum fidelity that can be achieved over all input
states.  This bound---and related ones that might come from using
similar theoretical techniques---should prove useful in analyzing the
performance of Gaussian channels.

\section*{Acknowledgments}
KW thanks P.~Grangier for interesting comments related to the mean
ensemble fidelity.




\end{document}